\documentclass{article}
\usepackage{psfig}
\usepackage{latexsym}
\makeatletter
\renewcommand{\thefootnote}{\fnsymbol{footnote}}

\newcommand{\be}{\begin{equation}}
\newcommand{\ee}{\end{equation}}
\newcommand{\ba}{\begin{eqnarray}}
\newcommand{\ea}{\end{eqnarray}}

\begin{document}

\begin{titlepage}
\vfill
\begin{flushright}
{\normalsize DAMTP-2002-155}\\
{\normalsize\tt hep-th/0212074}

\end{flushright}

\vskip 1in
\begin{center}
{\Large\bf Plane Waves and Vacuum Interpolation
}

\vskip 0.3in

{\large
Guillermo A. Silva\footnote{\tt G.A.Silva@damtp.cam.ac.uk}
}

\vskip 0.15in

{\it Department of Applied Mathematics and Theoretical Physics}\\
{\it University of Cambridge}\\ [3pt]
{\it Wilberforce Road, Cambridge, CB3 0WA,  U.K.} \\


\end{center}


\begin{abstract}
\normalsize\noindent 
A $\frac 12$-BPS family of time dependent plane wave spacetimes 
which give rise to exactly solvable string backgrounds is presented.
In particular a solution which interpolates 
between Minkowski spacetime and the maximally supersymmetric 
homogeneous plane wave along a timelike 
direction is analyzed. We work in  $d=4$, $N=2$ supergravity, 
but the results can be easily extended to $d=10,11$.
The conformal boundary of a particular class of solutions is 
studied.

\end{abstract}

\vfill

\end{titlepage}
\setcounter{footnote}{0}

\pagebreak
\renewcommand{\thepage}{\arabic{page}}
\renewcommand{\thefootnote}{\arabic{footnote}}
\pagebreak


\section*{Introduction}
Since the discovery of the maximally supersymmetric plane wave 
solution in type IIB theory \cite{bp}, 
great progress has been achieved in the AdS/CFT context in
going beyond the supergravity approximation \cite{str}. 
It is known that pp-waves 
yield exact classical backgrounds for string theory since all 
curvature invariants vanish and therefore receive no $\alpha'$ 
corrections. Hence, pp-wave spacetimes correspond to exact 
conformal field theories. This fact has been known for some 
time \cite{ht}, and after the work of \cite{bmn} the interest 
in them has been revived. 
Plane waves in particular provide simple toy models for studying 
string propagation and singularity issues because it has been shown
that they lead to exactly solvable models in the light cone gauge 
\cite{mets} (see also \cite{mm}).

After reviewing how to obtain the maximally supersymmetric solutions 
of $d=4$, $N=2$ supergravity in sections 1 
and 2, we present the $\frac12$-BPS 
time dependent pp-wave backgrounds in section 3. In section 4 we develop
a general method to study the conformal boundary of these time
dependent solutions and then for a particular class of 
metrics the boundary is analyzed. We end with some conclusions 
and directions for future research.

\section{Constraints from Maximal Supersymmetry }

We first review the spacetimes of $d=4$, $N=2$ supergravity with 
maximal supersymmetry, this means spacetimes with 8 
Killing spinors. The Killing spinor equation is 
\cite{gh}\footnote{See appendix for 
conventions.}
\be
\hat\nabla \xi\equiv(d+\frac 14 \omega-\frac i4 \not\!F\Gamma)\xi=0, 
\label{kil}
\ee
where $d=\partial_\mu\,dx^\mu$, 
$\omega=\omega_\mu^{~ab}\Gamma_{ab}\,dx^\mu$,
$\not\!\!F=F^{ab}\Gamma_{ab}$ and $\Gamma=\Gamma_a
\,e^a_{~\mu}dx^\mu$. Maximal supersymmetry implies that the
commutator of covariant derivatives as a spinor matrix must vanish
\be
\left[\hat\nabla_\mu,\hat\nabla_\nu\right]\xi=0,~\vee\!\!\!\!\!\!-\xi
\Rightarrow
\left[\hat\nabla_\mu,\hat\nabla_\nu\right]_{ss'}=0.
\label{comm}
\ee
Writing $\hat\nabla_\mu=D_\mu-\frac i4  \not\!F\Gamma_\mu$ where
$D_\mu\equiv\partial_\mu+\frac 14\omega_\mu^{~ab}\Gamma_{ab}$ the
commutator (\ref{comm}) takes the form
\be
\frac 14\Gamma_{ab} R^{ab}_{~~\mu\nu}-\frac i4
D_\mu( \not\!F\Gamma_\nu)+\frac i4 D_\nu( \not\!F\Gamma_\mu)
-\frac 1{16} \left[ \not\!F\Gamma_\mu,\not\!F\Gamma_\nu\right]=0.
\label{1}
\ee
Expanding (\ref{1}) in gamma matrices 
and setting to zero each independent coefficient 
we get for 
$\Gamma_{ab},\Gamma^\alpha,\Gamma^5\Gamma^\alpha$ and $\Gamma^5$
\be
R^{ab}_{~~\mu\nu}-e_{c\mu}\,e_{d\nu} (F^{ac}F^{bd}-F^{bc}F^{ad}+\tilde
F^{ac}\tilde F^{bd}-\tilde F^{bc}\tilde F^{ad})=0
\label{rie}
\ee
\be
D_\mu F_{\nu\alpha}-D_\nu F_{\mu\alpha}=0,
\label{f1}
\ee
\be
D_\mu \tilde F_{\nu\alpha}
-D_\nu \tilde F_{\mu\alpha}=0,
\label{f2}
\ee
\be
\tilde F_{\mu\rho}F^\rho_{~\nu}-\tilde F_{\nu\rho}F^\rho_{~\mu}=0,
\label{kg}
\ee
where $\tilde F_{\mu\nu}\equiv
\frac12\varepsilon_{\mu\nu\alpha\beta}F^{\alpha\beta}$. 
The set of equations (\ref{rie})-(\ref{kg}) are the constraints that 
a maximally supersymmetric spacetime of $d=4$, $N=2$ supergravity must 
satisfy \cite{kg}.

Equation (\ref{kg}) doesn't impose any 
constraint (cf. eqn.(8) of \cite{kg}), it is trivially satisfied 
due to the identity $F_{\mu}^{~\rho}\tilde F_\rho^{~\nu}=
-\frac14\delta_\mu^{\nu} F_{\rho\sigma}
\tilde F^{\rho\sigma}$.

Equations\ (\ref{f1}) and (\ref{f2}) imply, upon contracting with the
covariant Levi-Civita tensor, the Bianchi identity and
the  equations of motion  for $F_{\mu\nu}$. In particular, a
strong constraint follows for the electromagnetic field, by
substituting (\ref{f1}) into the Bianchi identity
\be
D_{\alpha}F_{\mu\nu}+D_{\mu}F_{\nu\alpha}+D_{\nu}F_{\alpha\mu}=0,
\ee
one finds that maximal supersymmetry requires that the
electromagnetic field must be covariantly constant
\be
D_{\alpha}F_{\mu\nu}=0.
\label{s1}
\ee
This is a natural property for a maximally supersymmetric
``vacuum'' solution.

The $d=4$ spaces admitting a covariantly constant 2-form were classified
in \cite{e}, where it was shown that there exist two
classes of solutions according to $F$ being either non-null
(Robinson-Bertotti \cite{rb}) or null 
(pp-waves \cite{exact},\cite{exact2}). One can
check that the Robinson-Bertotti $AdS_2\times S^2$ solution 
satisfies the 
constraint (\ref{rie}) as is well known \cite{garygift}. 
So let's move to the null electromagnetic field class of 
solutions.

\section{Maximally Supersymmetric Plane Waves}

To find the maximally supersymmetric plane wave solution \cite{kg}, 
we start from the general pp-wave solution of the Einstein-Maxwell 
action \cite{exact}
\ba ds^2&=&du(dv+H(u,\zeta,\bar\zeta)du)+d\zeta d\bar\zeta,\label{pp}\\
F&=&\frac 1{2}\,du\wedge(\partial_\zeta\phi\, d\zeta
+\partial_{\bar\zeta}\bar\phi\, d\bar\zeta),
\label{f}
\ea
where
\be
 H(u,\zeta,\bar\zeta)=f(u,\zeta)+\bar f(u,\bar\zeta)-\phi(u,\zeta)
 \,\bar\phi(u,\bar\zeta),~~~~
 F\equiv \frac 12 F_{\mu\nu} dx^\mu\wedge dx^\nu.
\ee
Here $f(u,\zeta)$ and $\phi(u,\zeta)$ are two arbitrary holomorphic
complex functions with complex conjugates $\bar f(u,\bar\zeta)$ 
and $\bar\phi(u,\bar\zeta)$.  Maximal supersymmetry constrains
the functional form of $f(u,\zeta)$ and
$\phi(u,\zeta)$.

Having a covariantly constant electromagnetic field (\ref{s1}) 
implies that
\be
 \phi(u,\zeta)=C(u)+{\lambda}\,\zeta.
 \label{fi}
\ee
The function $C(u)$ can be absorbed in $f(u,\zeta)$, and $\lambda$ can
always be made real by a rotation in $\zeta$, so we get
\ba
 H(u,\zeta,\bar\zeta)&=&f(u,\zeta)+\bar
 f(u,\bar\zeta)-\lambda^2\zeta\bar\zeta,\\
 F&=&\frac \lambda {2}\,du\wedge( d\zeta+d\bar\zeta).
\ea
The only independent non-zero components of the Riemann tensor for the
spacetime (\ref{pp}) are
\be
R_{uiuj}=-\frac 12\partial_i\partial_j H(u,\zeta,\bar\zeta),
\ee
where $i,j=(\zeta,\bar\zeta)$ refer to transverse space coordinates.

Equation (\ref{rie}) implies
\be
 \partial_\zeta\partial_{\bar\zeta}H=-\lambda^2,
\ee
\be
 \partial_\zeta\partial_\zeta H=
 \partial_{\bar\zeta}\partial_{\bar\zeta}H=0.\label{a}
\ee
The first equation is automatically satisfied because it is just the
Einstein equation\footnote{One can see that the
contraction of the Riemann tensor (\ref{rie}) gives 
Einstein equations
\be
R_{\mu\nu}=2(F_{\mu\alpha}F_{\nu\beta}g^{\alpha\beta}
-\frac 14 g_{\mu\nu}F^2).
\label {ein}
\ee}. Equations (\ref{a}) imply
that the space should be conformally flat and then
\be
 f(u,\zeta)=A(u)+B(u)\zeta.
 \label{efe}
\ee
Here $A(u),B(u)$ are two arbitrary complex functions of the real
variable $u$, but $f(u,\zeta)$ can be set to zero by a
diffeomorphism. 
Eqn. (\ref{efe}) together with (\ref{fi}) imply that 
the maximally supersymmetric pp-wave solution belongs 
to the subclass of plane wave
spacetimes. 
The maximally supersymmetric plane wave solution is  then
\be
 ds^2=du(dv-\lambda^2\zeta\bar\zeta\,du)+d\zeta d\bar\zeta,~~~~
 F=\frac \lambda 2\,du\wedge (d\zeta+d\bar\zeta).
 \label{compl}
\ee
This is called the Brinkmann form of the plane wave \cite{penrose}. 
Despite its appearence the spacetime (\ref{compl}) is completely 
homogeneous, having a seven dimensional group of isometries 
\cite{nw}\footnote{The number comes as follows: to the usual
$2d-3$ killing vectors that any plane wave spacetime 
has \cite{bfg}, in our case
5, one must add, from $H(u,\zeta,\bar\zeta)$ being 
$H(u,\zeta,\bar\zeta)=-\lambda^2\bar\zeta\zeta$, 
translations in $u$
and rotations in the transverse 
space $i(\zeta\partial_\zeta-\bar\zeta\partial_{\bar\zeta})$. 
The appearance of $\partial_u$ as a Killing vector makes the 
spacetime homogeneous.
}. 
The isometry group of $({\cal M},F)$ is six dimensional
and this coincides with the dimension of the isometry group of the 
Robinson-Bertotti solution. This observation was at the root 
of the connection between the homogeneous plane waves and 
the $AdS\times S$ via Penrose limits and group contractions 
\cite{pl}\cite{bmn}\cite{hks}.

We have seen that maximal supersymmetry for $d=4$, $N=2$
supergravity implies 
that the spacetime must be 
conformally flat. This statement extends to $d=6,10$ but not
to $d=5,11$ \cite{bfg},\cite{mes}.

\section{$\frac 12$-BPS Time Dependent Plane Wave Solutions}

Now we are going to find some new time dependent $\frac 12$-BPS
solutions which can be used in $d=10$ as string backgrounds 
leading to gaussian models in the light-cone gauge. 

The Killing spinor equations (\ref{kil}) for the background
(\ref{pp})-(\ref{f}) take the form
\ba
 \left(\partial_\zeta-\frac i {8}\partial_\zeta\phi\, 
 \Omega\right)\xi&=&0,\label{sp1}\\
 \left(\partial_{\bar\zeta}-\frac i {8}{{\partial_{\bar\zeta}
 \bar\phi}}\, \Omega'\right)\xi&=&0,\label{sp2}\\
 \left(\partial_u+\frac14 \Upsilon\right)\xi&=&0,\label{sp3}\\
 \partial_v\,\xi&=&0,\label{sp4}
\ea
where 
\ba
 \Omega&=&\Gamma^1\Gamma^2\Gamma^-,\label{ome}\\ 
 \Omega'&=&\Gamma^2\Gamma^1\Gamma^-,\label{ome'}\\ 
 \Upsilon&=&\Gamma^-(\partial_\zeta H\,\Gamma^1
 +\partial_{\bar\zeta} H\,\Gamma^2)+\frac i {2}(\partial_\zeta 
 \phi\,
 \Gamma^1+{\partial_{\bar\zeta}\bar\phi}\, 
 \Gamma^2)\Gamma^-\Gamma^+.
\ea
The spinor projection\footnote{The coodinate independent
way of expressing (\ref{proj}) is $(l\cdot\Gamma)\,\xi=0$ where $l$
is the covariantly constant null vector of the pp-wave \cite{h}.}
\be
 \Gamma^-\xi=0,\label{constr}
 \label{proj}
\ee
was used in  \cite{gh},\cite{lu}. We now further
impose the constraint 
\be
 \phi(u,\zeta)=g(u)\,\zeta,
 \label{ans}
\ee
where $g(u)$ is an arbitrary complex function 
and show that we have $\frac12$-supersymmetry. Using the
constraint (\ref{proj}), the system  
(\ref{sp1})-(\ref{sp4}) reduces to
\be
 \partial_v\xi=\partial_\zeta\xi=\partial_{\bar\zeta}\xi=0,
\ee
\be
 \left(\partial_u+\frac i2(\partial_\zeta\phi\,\Gamma^1+
 \partial_{\bar\zeta}\bar\phi\,\Gamma^2)\right)\xi=0,
\label{k}
\ee
and we see that a consistent $\frac12$-BPS solution is found if 
(\ref{ans}) is satisfied. 

From now on we  take $g(u)$ to be real. 
The solution for the Killing spinors is 
\be
 \xi=e^{-\frac {i}{2}(\Gamma^1+\Gamma^2)\,{\cal G}(u)}\xi_0,
 \label{1/2}
\ee
where ${\cal G}(u)=\int_{u_0}^udu'g(u')$ and $\xi_0$ satisfies (\ref{constr}).
The exponential in (\ref{1/2}) can be resummed using the
pa\-ra\-me\-tri\-za\-tion for the gamma matrices given in the appendix,
obtaining
\be
 \xi=\left\{\cos ({\cal G}(u))
 +\gamma^2\sin ({\cal G}(u))\right\}\xi_0.
\ee
This shows that the Killing spinors obtained by the proyection 
(\ref{constr}) always depend on the ``time-like'' $u$
coordinate, are independent of the transverse coordinates
and correspond to kinematical supersymmetries 
\cite{mt}. 

For the maximally supersymmetric 
solutions $g(u)$ is constant, ${\cal G}(u)$ becomes linear and then 
half of the spinors are periodic in the $u$ coordinate. Explicitly,
the Killing spinors are given by
\be
 \xi=(1+\frac {i\lambda}8 (\Omega\zeta+\Omega'\bar\zeta))
 e^{-\frac {i\lambda u}8(\Gamma^1+\Gamma^2)\Gamma^-\Gamma^+}\xi_0.
\ee
Using the parametrization given in the appendix, the exponential
can be resummed obtaining
\be
 \xi=(1+\frac {i\lambda}8 (\Omega\zeta+\Omega'\bar\zeta))
 (A\cos \lambda u-iB\sin\lambda u)\xi_0,
 \label{expl}
\ee
(expressions for $A$ and $B$ are given in the appendix). Eqn. 
(\ref{expl}) shows the periodicity of the spinors in
the $u$ coordinate and from the eigenvalues of $A$ and $B$ 
one gets that only half of the spinors depend explicitly on $u$.

\section{Conformal Boundary}

From the discussion of the last section we know that a 
$\frac 12$-BPS family of solutions is
obtained by the ansatz (\ref{ans})
\ba 
 ds^2&=&dudv-g^2(u)\, x_i^2\,du^2+d x_i^2,\label{interg}\\
 F&=& {g(u)}\,du\wedge dx_1,
 \label{interf}
\ea
here $i=1,2$ refers to the {\it real} transverse directions. 
Computing the curvature of (\ref{pp}), one notices that the
only nontrivial pp-waves which are non-singular are
precisely of the form (\ref{interg}) with smooth $g(u)$ \cite{hs}. 
In the $g(u)=const.$ case, supersymmetry gets enhanced
and we get the fully supersymmetric solution (\ref{compl}).

From the form (\ref{interg}) we see that the spacetime is 
foliated by null hypersurfaces $u=const.$ and the vector 
$t=\partial_u$ connecting these is timelike everywhere except
at the origin of the transverse space.

We will now analyse the conformal boundary of the solution
(\ref{interg}) for particular choices of $g(u)$. The plan is,
given that the solutions are conformally flat,
to embed them into the Einstein Static Universe (ESU) and study 
where the conformal factor blows up. This is done 
by going from Brinkman (\ref{interg}) to Rosen coordinates \cite{gary75}
where the metric takes the form\footnote{The change of coordinates to 
go from (\ref{interf}) to (\ref{pprosen}) is $v=\tilde v-p(u)\dot
p(u)\,\tilde x_i^2,~~x_i=p(u)\,\tilde x_i$ where the index 
$i=1,2$ refers to the 
transverse space coordinates. $p(u)$ is a solution of
(\ref{ed}) which can be interpreted as a harmonic oscillator
with time dependent frequency.}
\be
 ds^2=dud\tilde v+p^2(u)\,d\tilde x_i^2,
 \label{pprosen}
\ee
here $p(u)$ is given by the 
solution of 
\be
 \ddot p(u)=-g^2(u)\,p(u).
 \label{ed}
\ee
Factorizing out $p(u)^2$
in (\ref{pprosen}) one obtains the conformally flat 
expression
\be
 ds^2=\frac 1{\Pi^2(\tilde u)}(d\tilde ud\tilde v+\,d\tilde x_i^2).
 \label{cf}
\ee
where $1/\Pi^2(\tilde u)=p^2(u(\tilde u))$ and 
$\tilde u=\int \frac 1{p^2(u)}du$. 
Now using the standard coordinate transformations 
\cite{he} 
\be
 \tilde u=\frac {\sin\psi+\sin\xi\cos\theta}{\cos\psi+\cos\xi},~~
 \tilde v=\frac {-\sin\psi+\sin\xi\cos\theta}{\cos\psi+\cos\xi},~~
 \tilde x_1=\frac{\sin\xi\sin\theta\cos\phi}{\cos\psi+\cos\xi},~~
 \tilde x_2=\frac{\sin\xi\sin\theta\sin\phi}{\cos\psi+\cos\xi},
 \label{esucoor}
\ee
one can embed (\ref{cf}) into the ESU. The result is
\be
 ds^2=\frac1{(\cos\psi+\cos\xi)^2\Pi^2(\tilde u)}
 (-d\psi^2+d\xi^2+\sin^2\!\xi\, (d\theta^2+\sin^2\theta\, d\phi^2)).
 \label{cesu}
\ee
The conformal boundary is defined as the set of points where  
the conformal factor 
\be
 \Omega(x^\mu)=(\cos\psi+\cos\xi)\,\Pi(\tilde u),
\ee
vanishes. The normal to the hypersurfaces $\Omega(x^\mu)=const.$ is
given by $n_\mu=\partial_\mu\Omega$ and the nature of the boundary
can be analysed by computing
\be
 n^2=\frac 12 \Pi(\tilde u)\left(\Pi(\tilde u)\,
 (\cos2\psi-\cos2\xi)+4\dot\Pi(\tilde u)\,(\sin\psi\cos \xi-
 \cos\psi\sin\xi\cos\theta)\right).
 \label{norm}
\ee
and evaluating it at $\Omega=0$. 

~

i) Minkowski space: This very well known case corresponds to 
$g(u)=0$ in (\ref{interg}). In  Rosen coordiantes we get 
$p(u)=A+Bu$, where $A,B$ are two arbitrary constants. By 
taking $B=0$ we end up in ESU with the standard expression 
\be
  ds^2=\frac{1}{(\cos\psi+\cos\xi)^2}
 (-d\psi^2+d\xi^2+\sin^2\!\xi\, (d\theta^2+\sin^2\theta\,d\phi^2)).
 \label{ma}
\ee
In this case $\Pi(\tilde u)=1$ and computing (\ref{norm}) on 
$\Omega=0$ we see that the boundary is a null co-dimension 1 
hypersurface.

A non standard expression for the conformal compactification
of Minkowski space is obtained by choosing $A=0$ and $B=1$,
this is
\be
  ds^2=\frac{1}{(\sin\psi+\sin\xi\cos\theta)^2}
 (-d\psi^2+d\xi^2+\sin^2\!\xi\, (d\theta^2+\sin^2\theta\,d\phi^2)).
 \label{mb}
\ee
Here we get $\Pi(\tilde u)=\tilde u$ and of course we obtain
again a co-dimension 1 null boundary

\begin{figure}
 \vspace{-1cm}
 \centerline{ \psfig{figure=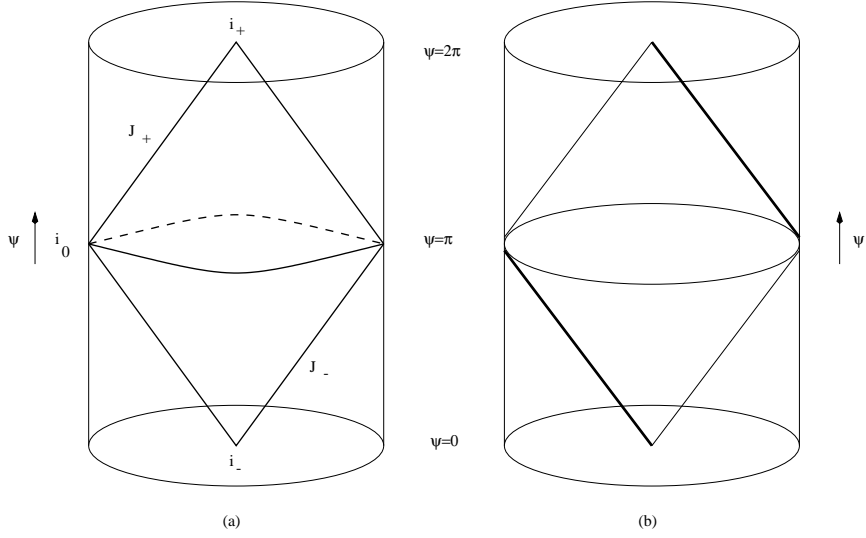,height=7cm,angle=0}}
 \caption{(a) Minkowski spacetime eqn.(\ref{ma}). (b) Hpw 
 spacetime eqn.(\ref{maxsus}). 
 The base of the solid cylinders represents $S^3$, 
 the center of it being $z_4=-1$ and the boundary being $z_4=1$. 
 The vertical direction corresponds to the time coordinate $\psi$. As 
 usual, the boundary of the cylinder should be regarded as 
 1 dimensional. The thick lines in (a) and (b) correspond to the set 
 of points where the conformal factors in (\ref{ma}) and 
 (\ref{maxsus}) vanish.  The time 
 coordinate difference between the tips of the cones is 
 $\Delta\psi=2\pi$.}
 \label{pene} 
\end{figure}

Picturing the $S^3$ parametrized by $(\xi,\theta,\phi)$ as 
$z_1^2+z_2^2+z_3^2+z_4^2=1$ 
where 
\be
 z_1=\sin\xi\sin\theta\cos\phi,
 ~z_2=\sin\xi\sin\theta\sin\phi,~z_3=\sin\xi\cos\theta,
 ~z_4=\cos\xi,
 \label{s3}
\ee
one sees that the conformal factors in (\ref{ma}) or (\ref{mb}) 
correspond to the slicing of the $S^3$ by $z_4=-\cos\psi$ or 
$z_3=-\sin\psi$ respectively. This gives rise to
the  pictorial representation of  compactified Minkowski 
space as the interior of a cone followed by an inverted cone 
(see figure \ref{pene}.(a)).

~

ii) Maximally supersymmetric homogeneous plane wave (Hpw): 
The conformal boundary 
for this spacetime in the $d=10$ case has been studied in 
\cite{bn} (see also \cite{mr}). Setting $g(u)=1$ by an 
appropiate change of coordinates, we obtain
\be
  ds^2=\frac{1}{(\cos\psi+\cos\xi)^2+(\sin\psi+\sin\xi\cos\theta)^2}
 (-d\psi^2+d\xi^2+\sin^2\!\xi\, (d\theta^2+\sin^2\theta\,d\phi^2)).
 \label{maxsus}
\ee
We get $\Pi(\tilde u)=\sqrt{1+\tilde u^2}$ and by  (\ref{norm}) 
we see again that 
the boundary is a null surface. In this example, however, the conformal
boundary is
1-dimensional and this comes about because we must look for the 
intersection of the two null hypersurfaces ${\cal
S}_1(x^\mu)=\cos\psi+\cos\xi=0$ and ${\cal S}_2(x^\mu)=\sin\psi+
\sin\xi\cos\theta=0$.
Na\"\i vely one would expect a co-dimension 2 surface (co-dimension $D-2$ in the 
general case), but in this
particular case the intersection of the two hypersurfaces 
${\cal S}_1(x^\mu)=\cos\psi+z_4=0$ and 
${\cal S}_2(x^\mu)=\sin\psi+z_3=0$ 
occurs for any value of $\psi$ at the north/south pole 
of the $S^2_{\theta,\phi}$,
where the fiber corresponding to the $\phi$ coordinate 
in (\ref{cesu}) collapses. In other words this means 
that at any given fixed time $\psi$
the intersection of the two null hypersurfaces is 
just one point. In the higher 
dimensional cases this same phenomenon happens and the conformal 
null boundary of the maximally supersymmetric plane waves
is again 1-dimensional (see figure \ref{pene}.(b)).

~

iii) Minkowski$\to$Hpw$\to$Minkowski plane wave (MHMpw): An analytically 
tractable example is obtained by choosing $g(u)$ of the form
\be
 g(u)=\frac 1{1+a^2u^2}.
 \label{ansa}
\ee
For large absolute values values of $u$ we asymptotically get
Minkowski spacetime and the homogeneous plane-wave 
region appears for  $u\approx 0$. In particular in the limit 
$a\to\infty$ we get Minkowski space and for
$a\to 0$ we get
the homogeneous plane wave (\ref{maxsus}). 

The solution to (\ref{ed}) is
\be
 p(u)=A_1\sqrt{1+a^2u^2}\,
 \cos\left(\frac {\sqrt{1+a^2}}a \arctan(au)+A_2\right).
\ee
Without loss of generality we can take $A_1=1$ and $A_2=0$, 
the expression for $\Pi(\tilde u)$ is\footnote{After an
obvious rescaling in $\tilde u$.}
\be
 \Pi^2(\tilde u)=\frac {1+\tilde u^2}{1+\tan^2
 \left(\frac 1\eta \arctan\tilde u \right)},
\ee
where $\eta=\frac {\sqrt{1+a^2}}a>1$. (Minkowski spacetime is obtained 
for $\eta=1$ and the maximally supersymmetric Hpw for $\eta=\infty$.) 
The embedding in ESU takes the form
\be
 ds^2=\frac {1+\tan^2
 \left(\frac 1\eta \arctan\left(\frac {\sin\psi+\sin\xi\cos\theta}
 {\cos\psi+\cos\xi}\right)\right)}
 {(\cos\psi+\cos\xi)^2+(\sin\psi+\sin\xi\cos\theta)^2}
 \,ds^2_{ESU}.
 \label{inter}
\ee
From this expression it is clear that the conformal boundary will
be given by the set of points corresponding to the 1-d ``helix'' 
described in the last subsection, 
where the denominator vanishes, plus the ones which make
the numerator of (\ref{inter}) blow up, this means the argument 
of the tangent  in the 
numerator of (\ref{inter}) being equal to $\frac\pi2+n\pi$.

\begin{figure}
 \vspace{-1cm}
 \centerline{ \psfig{figure=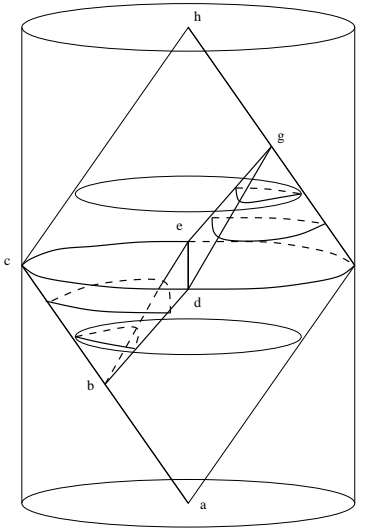,height=7cm,angle=0}}
 \caption{Visualization of the plane front $\tilde u=0$ inside
 Minkowski spacetime (cf. figure 1.(a)). The null hypersurface consists
 of three parts:  between the points `a'  and `b', and the points 
 `g' and `h' it is 1-dimensional and just coincides with the 1-d {\it
 helix} found in \cite{bn}, between points `b' and `g' the hypersurface is
 3-dimensional. The `cones' originated from `b' and `g' remain
 strictly inside Minkowski spacetime coinciding with the conformal
 boundary of it
 only along the 1-d {\it helix} (ac + fh). The points `c', `d', `e' and `f' 
 should be identified.}
 \label{plf} 
\end{figure}

To understand the conformal boundary of (\ref{inter}) 
it is necessary to understand how the $\tilde u=const.$ 
null hypersurfaces in (\ref{cf}) get 
compactified when we go go to the ESU 
(recall that the argument of $\arctan$ is just
$\tilde u$ see eqn.(\ref{esucoor})). 
It is important to have in mind that these null 
hypersurfaces 
which we will take to be $\tilde u=t+\tilde x_3$ correspond to
plane fronts and should be distinguished from the ones in Penrose's 
compactification which are spherical fronts $\tilde u=t+r$. By 
looking at the hypersurfaces as $\psi$-dependent intersections of the 
$S^3$ parametrized by $(\xi,\theta,\phi)$, from (\ref{esucoor}) and 
(\ref{s3}) we get that they correspond to the slicing
\be
  z_3=\tan\alpha\, z_4+\frac {\sin(\alpha-\psi)}{\cos \alpha}
 \label{sli}
\ee
where $\tilde u=\tan\alpha$. A visualization
for the $\tilde u=0$ case is given in figure \ref{plf}. In the
general case the null hypersurface consists of three parts, two
1-dimensional sectors represented in figure \ref{plf} by the 
lines `ab' and `gh'  and one 3-dimensional between points `b' 
and `g'. As $\tilde u$ becomes negative the points `b' and `g' 
approach `a' and `f' respectively, and the points `d' and `e' get
closer to `f'. In the case of $\tilde u$ becoming positive  all
the points mentioned before move in opposite direction. 
With this picture in mind one concludes that the top and bottom cones
in the figure \ref{pene}.(a) should be identified with the 
$\tilde u=\pm\infty$ respectively.
Furthermore, from this viewpoint one recognizes the 
conformal boundary of the Hpw as nothing but the intersection 
of the boundary of Minkowski spacetime (figure 1.(a)) with the  
$\tilde u=const<\infty$ null hyperplane (cf. with the definitions
in terms of ${\cal S}_{1,2}$ in (ii)).

\begin{figure}
 \vspace{-1cm}
 \centerline{ \psfig{figure=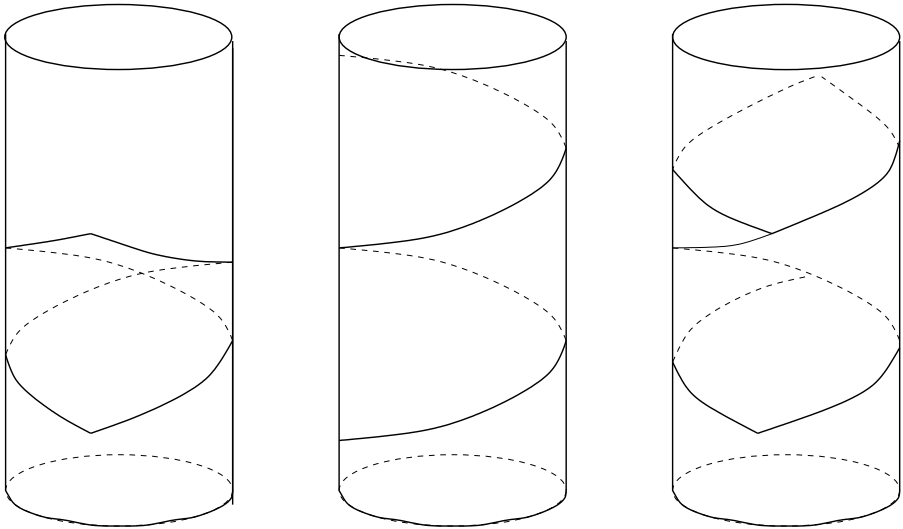,height=7cm,angle=0}}
 \caption{ Embedding of Minkowski, Hpw and MHMpw spacetimes
 into the Einstein Static Universe. The ESU is pictured as the 
 surface of the cylinder, the circle at constant $\psi$ represent the
 3-sphere $S^3$. }
 \label{penemb} 
\end{figure}

Returning to (\ref{inter}) one immediately realizes that for $\eta>1$ 
there exists no value of $\tilde u\in (-\infty,\infty)$ making the 
numerator blow up. However, an important point is worth mentioning: 
when going from (\ref{interg}) to the conformally embedded 
expression (\ref{cesu}) taking (\ref{ansa}), one can 
see  that the coordinate range 
$\tilde u\in(-\infty,\infty)$, used as an intermediate step 
to obtain (\ref{inter}), only covers the patch 
$|u|<\frac 1a\tan(\frac {a}{\sqrt{1+a^2}}\frac \pi2)$ of the 
original Brinkman coordinate system. This fact is manifested 
mathematically in (\ref{inter}) by the presence of the 
$\arctan$ function which should then be analytically 
continued in order to cover the original geodesically 
complete spacetime. To be precise: for 
$\psi\in(2n\pi,(2n+2)\pi)$ the value of $\arctan$ should be
restricted to be within $((-\frac 1 2+n)\pi,(\frac 12+n)\pi)$.
With this prescription it becomes clear that the conformal
boundary of the spacetime (\ref{inter}) will consist of
two null  $\tilde u=const$ hypersurfaces (see 
figure \ref{plf}) where the numerator
of (\ref{inter}) will blow up,  joined by the 1-dimensional 
null line depicted in figure \ref{pene}.(b), where the 
denominator in (\ref{inter}) vanishes. One may say that the
two cones present in the conformal compactification of
Minkowski space and intersecting at $i_0$ have now been
separated and joined by the null 1-dimensinal {\it helix}
(see figure \ref{penemb}). The $\psi$ 
separation between the two null hypersurfaces is related
to the value of $\eta$ in (\ref{inter}). As $\eta$ becomes
bigger and bigger, the two null hypersurfaces become more
and more separated. The opposite behaviour occurs when
$\eta$ approaches 1.

\section*{Conclusion and Discussion} 

We have obtained new $\frac12$-BPS time dependent backgrounds 
which when extended to $d=10$ will give  gaussian models
with ``time dependent masses'' in the light-cone gauge.

A general procedure for analyzing the conformal boundary of
the time dependent spacetimes was developed and applied to
a particular example.

It may perhaps be worth noticing that for the particular
class of spacetimes (\ref{interg}) the only non-zero
Ricci component is $R_{uu}$, this means that they can be 
supported by any p-form of the form 
$F=g(u)\,du\wedge \varphi$ where $\varphi$ is a constant 
(p-1)-form in the transverse flat space. The resulting
background can then be choosen to be soported by 
Neveu-Schwarz or Ramond-Ramond field with 
constant dilaton field. For the particular
case of $g(u)=const$ the background supported by the
NS 3-form field strength was obtained long ago by
a WZW construction over a generalized Heisenberg
group \cite{km}.

The study of strings propagating on these  
backgrounds is under investigation.

~

\underline{N.B}: While this work was being written up, the paper
\cite{trp} appeared where similar issues are discussed.

\vskip 1cm

\noindent
\leftline{\Large \bf Acknowledgments}
\vskip 0.5cm \noindent
It is pleasure to thank Pascal Bain, 
Sean Hartnoll, Venkata Suryanarayana Nemani,  Carlos N\'u\~nez,
Rub\'en Portugu\'es and especially Gary Gibbons for many 
fruitful conversations. Research supported by  
CONICET  and Fundaci\'on Antorchas, Argentina.

\vskip 1cm
\appendix

\section*{Appendix}

\subsection*{General pp-wave solution}

The covariantly constant null vector of (\ref{pp}) is
$l=l^\mu\partial_\mu=\partial/\partial v$ and  we have used
light-cone coordinates $(u,v)$ and complex coordinates
$(\zeta,\bar\zeta)$ related to the usual cartesian ones by
$\zeta=x_1+i\,x_2$. The Riemann tensor for the spacetime (\ref{pp})  
can be written as
$R_{\mu\nu\rho\sigma}=l_{[\mu}k_{\nu][\rho}l_{\sigma]}$ where
$l_\mu$ are the covariant components of the null constant vector
and the symmetric tensor $k_{\mu\nu}$ is non-zero only for the
transverse coordinates $i,j=\zeta,\bar\zeta$ and has the form
$k_{ij}=\frac 12 \partial_i\partial_j H$. The electromagnetic field
(\ref{f}) can be
written as $F_{\mu\nu}=l_{[\mu}s_{\nu]}$ where 
$s_\mu=\sqrt2
(\partial_{\zeta}\phi,\partial_{\bar\zeta}\bar\phi,0,0)$ in 
the $x^\mu=(\zeta,\bar\zeta,u,v)$ coordinate system. Note that
$l\cdot s=0$.

The metric (\ref{compl}) can also be  obtained from a Wess-Zumino-Witten
construction over the Heisenberg-Cangemi-Jackiw (HCJ) group \cite{nw} 
(the $d=6,10$ maximally supersymmetric plane waves can be obtained 
as well as the bi-invariant
metrics over generalized HCJ-groups 
via a WZW construction \cite{km}).
The expresion found by Nappi-Witten is obtained by making in 
(\ref{compl})  the change of 
coordinates $\zeta=e^{-i\lambda u}\omega$
\be
  ds^2=dudv+i\lambda(\bar\omega d\omega-\omega d\bar\omega) du+d\omega
  d\bar\omega,~~~~F=\frac \lambda{2}
 du\wedge(e^{-iCu}d\omega+e^{iCu}d\bar\omega).
\ee
In cartesian coordinates $\omega=y_1+i\,y_2$ this takes the form 
$$
 ds^2=dudv+2\lambda(y_2 dy_1-y_1 dy_2) du+dy_1^2+dy_2^2,~~~~
 F=\lambda\, du\wedge(\cos(\lambda u)dy_1+\sin(\lambda u)dy_2)
$$
Note that the maximally supersymmetric solution is 
supported by a 1-form gauge field
instead of a 2-form gauge fields. The $d=10$ case works along
the same lines, with the metric supported by the self-dual 
RR 5-form and having constant dilaton.

\subsection*{Tetrads, spin connections and gamma matrices conventions}

 My conventions for $\Gamma$
matrices are: $\{\Gamma_a, \Gamma_b\}=2\eta_{ab}$ with $\eta_{ab}$
mostly plus,
$\Gamma_{ab}=\frac12\left[\Gamma_a,\Gamma_b\right]$,
$\Gamma^5=\frac i{4!}\epsilon_{abcd}\Gamma^a\Gamma^b\Gamma^c\Gamma^d$ and
$\epsilon_{0123}=-1$. $a,b\ldots$ are flat indices while
$\mu,\nu\ldots$ are spacetime indices.

For the general pp-wave solution (\ref{pp}) I choose the 
tetrads
\be
 e^{ 1}=d\zeta\,,~~~~e^{ 2}=d\bar\zeta\,,~~~~
 e^{-}=du\,,~~~~e^{+}=dv+H(u,\zeta,\bar\zeta)\,du,
\ee
which have the tangent space inverse metric  
\be
 \eta^{ab}=\left(
 \begin{array}{lrrr}
    0 & 2 & 0 & 0\\
    2 & 0 & 0 & 0\\
    0 & 0 & 0 & 2 \\
    0 & 0 & 2 & 0
 \end{array} \right).
\ee
Here $a,b=1,2,-,+$. The spin connections are given by
\be
 \omega_{ {-1}}=\frac12 \partial_\zeta H\,du~,~~~~~
 \omega_{ {-2}}=\frac12 \partial_{\bar\zeta} H\,du.
\ee
A real representation of the gamma matrices satisfying
$\left\{\Gamma^a,\Gamma^b\right\}=2\eta^{ab}$ is given by
\be
 \Gamma^1=i\gamma^2+\gamma^3,~~~
 \Gamma^2=i\gamma^2-\gamma^3,~~~
 \Gamma^-=\gamma^0+\gamma^1,~~~
 \Gamma^+=\gamma^0-\gamma^1,
\ee
where $\gamma^\mu$ are the standard chiral representation 
for the gammas with mostly minus flat 
metric given by
\be
 \gamma^{\mu}=\left(
 \begin{array}{lr}
    0 & \sigma^\mu \\
    \bar\sigma^\mu & 0
  \end{array} \right),
\ee
$\sigma^\mu=(I_2,\vec \sigma),~\bar\sigma^\mu=(I_2,-\vec
\sigma)$. The relation between curved and tangent space  gamma matrices
$\Gamma_\mu=e_{a\mu}\Gamma^a$ gives
\be
 \Gamma_\zeta=\frac12 \Gamma^2,~~~\Gamma_{\bar\zeta}=\frac12 \Gamma^1
 ,~~~\Gamma_u=\frac12 (\Gamma^++H\, \Gamma^-),~~~
 \Gamma_v=\frac12 \Gamma^-
\ee
The matrices (\ref{ome})-(\ref{ome'}) take the explicit form
\be
 \Omega=\left(
 \begin{array}{lr}
    ~~~~~~0 & 0 \\
    4(I_2-\sigma^1) & 0
  \end{array} \right),~~~
 \Omega'=\left(
 \begin{array}{lr}
    0 & 4(I_2+\sigma^1) \\
    0  & 0~~~~~~
  \end{array} \right).
\ee
The matrices $A$ and $B$ appearing in (\ref{expl}) are
\be
 A=\frac 12\left(
 \begin{array}{lr}
    I_2+\sigma^1 & 0~~~~ \\
    ~~~~0  &  I_2-\sigma^1
  \end{array} \right),~~~
 B=\frac 12\left(
 \begin{array}{lr}
    ~~~~~0 & -\sigma^3+i\sigma^2 \\
    -\sigma^3-i\sigma^2 & 0~~~~~
  \end{array} \right).
\ee
Both $A,B$ are hermitian and commute between 
themselves. $A$ has eigenvalues $\lambda_A=0,0,1,1$ whereas 
$B$ has  $\lambda_B=0,0,1,-1$. In 
other words, $A$ and $B$ have simultaneously
4 null eigenvectors.


\newcommand{\J}[4]{{\sl #1} {\bf #2} (#3) #4}
\newcommand{\andJ}[3]{{\bf #1} (#2) #3}
\newcommand{\AP}{Ann.\ Phys.\ (N.Y.)}
\newcommand{\MPL}{Mod.\ Phys.\ Lett.}
\newcommand{\NP}{Nucl.\ Phys.}
\newcommand{\PL}{Phys.\ Lett.}
\newcommand{\PR}{ Phys.\ Rev.}
\newcommand{\PRL}{Phys.\ Rev.\ Lett.}
\newcommand{\PTP}{Prog.\ Theor.\ Phys.}
\newcommand{\hep}[1]{{\tt hep-th/{#1}}}

\end{document}